\documentstyle[12pt,epsf,amstex]{article}

\begin{document}
\setlength{\unitlength}{1mm}
\textwidth 15.0 true cm 

\headheight 0 cm
\headsep 0 cm
\topmargin 0.4 true in
\oddsidemargin 0.25 true in
\input epsf.sty

\newcommand{\beq}{\begin{equation}}
\newcommand{\eeq}{\end{equation}}
\newcommand{\be}{\begin{eqnarray}}
\newcommand{\ee}{\end{eqnarray}}
\renewcommand{\vec}[1]{{\bf #1}}
\newcommand{\vecg}[1]{\mbox{\boldmath $#1$}}

\renewcommand{\theequation}{\thesection.\arabic{equation}}

\newcommand{\grpicture}[1]
{
    \begin{center}
        \epsfxsize=200pt
        \epsfysize=0pt
        \vspace{-5mm}
        \parbox{\epsfxsize}{\epsffile{#1.eps}}
        \vspace{5mm}
    \end{center}
}

\begin{flushright}

SUBATECH--02--20\\

\end{flushright}

\vspace{0.5cm}

\begin{center}

{\Large\bf  
Abelian Matrix Models in Two Loops.}

\bigskip

   {\Large  A.V. Smilga} \\

\vspace{0.8cm}

{\it SUBATECH, Universit\'e de
Nantes,  4 rue Alfred Kastler, BP 20722, Nantes  44307, France. }\\

\end{center}

\bigskip

\begin{abstract}
We perform a two-loop calculation of the effective Lagrangian for
 the 
low--energy modes of the quantum mechanical system obtained by 
dimensional
reduction from 4D, ${\cal N} = 1$ supersymmetric QED. The bosonic 
part of
the Lagrangian describes the motion over  moduli space of vector 
potentials
$A_i$ endowed with a nontrivial conformally flat metric
$$ g_{ij} \ =\ \delta_{ij} \left( 1 + \frac 1{2|\vec{A}|^3} - 
\frac 3{4|\vec{A}|^6} + \ldots \right)\ . $$
For the matrix model obtained from Abelian 4D,  ${\cal N} = 2$ 
theory, the
two--loop correction $\propto 1/|\vec{A}|^6 $ vanishes as it 
should. 
\end{abstract}

\section{Introduction}
\setcounter{equation}0

Back in 1987 we determined the one--loop corrections to the 
effective
Born--Oppenheimer Hamiltonian of 4D, 
${\cal N} = 1$ 
supersymmetric QED  reduced to $(0+1)$ dimensions \cite{jaSQED}. 
The corresponding effective 
Lagrangian
is expressed in terms of the real supervariable carrying vector 
index $k$ 
\cite{Ivanov}, 
  \be
 \label{phiAB}
\Phi_k \ =\  -\frac 14 \epsilon^{\beta\gamma}
(\sigma_k)_\gamma^{\ \alpha} (D_\alpha \bar D_\beta + D_\beta 
\bar D_\alpha) V\ ,
 \ee
where $D_\alpha, \bar D_\alpha$ are supersymmetric covariant 
derivatives and V is a scalar real supervariable, the quantum--mechanical
descendant of the $4D$ vector superfield. The lowest
component of $\Phi_k$ is the vector potential $A_k$. 
Remarkably, $A_k$ and $\Phi_k$ are gauge invariant in the QM limit. 
A generic supersymmetric
Lagrangian involving $\Phi_k$ is
 \be
\label{LFPhi}
L\ =\ \int  d^2\theta d^2 \bar\theta \ F(\vecg{\Phi})\ ,
 \ee
Its bosonic part  describes the motion along a 
3--dimensional manifold with the conformally flat metric 
 \be
\label{metric}
ds^2 \ =\ 2  \Delta F(\vec{A}) d\vec{A}^2\ .
 \ee
Explicit one-loop calculations give the result
\footnote{This analysis has been extended to non-Abelian case in 
\cite{BOnab}.
For example, for $SU(2)$ ${\cal N} = 1$ theory, the effective 
Lagrangian is
given, again, by Eq.(\ref{LFPhi}) with 
$$ 2 \Delta F(\vec{A}) \ =\ 1 - \frac 3{2|\vec{A}|^3} + \ldots $$
 }
 \be
\label{metr1}
2 \Delta F(\vec{A}) \ =\ 1 + \frac 1{2|\vec{A}|^3} + \ldots
 \ee
We noted back in \cite{jaSQED} that the coefficient of 
$1/|\vec{A}|^3$
in Eq.(\ref{metr1}) is related to the first coefficient of the
4--dimensional $\beta$ function of supersymmetric QED. In recent 
\cite{Akh}
we reproduced the result (\ref{metr1}), obtained originally in the 
framework
of the Hamiltonian Born--Oppenheimer procedure, by Lagrangian 
methods and showed
that the term $\propto 1/|\vec{A}|^3$ is determined by exactly the 
same graph
as the 1--loop correction to the  effective Lagrangian in 
4 dimensions.

The effective Lagrangia both in (0+1) and in (3+1) theories accept 
also
higher loop corrections. It is natural to expect that they are 
related
to each other, as the one--loop corrections do. In particular, 
higher-order
corrections vanish both in ${\cal N} =2$ (3+1) theories and in 
their 
quantum-mechanical counterparts. It is unconceivable for us that 
this is
a purely accidental coincidence \cite{Akh}. 

However, to see a relationship between the corrections at the 
two--loop
level or higher is not an easy task. At the one--loop level the 
basic rule of 
correspondence is
 \be
\label{corresp}
\frac 1 { 4|\vec{A}|^3} = \int_{-\infty}^\infty \frac 
{d\omega}{2\pi}
\frac 1{(\omega^2 + \vec{A}^2)^2} \ \longrightarrow \ 
\int \frac {d^4p}{(2\pi)^4}
\frac 1{(p^2 + \mu^2)^2}   = \frac 1{16\pi^2} \ln \frac 
{\Lambda_{UV}^2}{e\mu^2} \ .
 \ee
The correspondence between multiple loop integrals in different 
dimensions
is much more obscur. The basic motivation of the present study was 
the
desire to establish such a correspondence. We have to say right 
away that we
{\it failed} to do it. The correction to 
${\cal L}_{\rm eff}^{(0+1)}$ turned
out to be
 \be
\label{corr01}
{\cal L}_{\rm eff}^{(0+1)} \ =\ \frac {\dot{\vec{A}}^2}2 
\left( 1 + \frac 1{2|\vec{A}|^3} - 
\frac 3{4|\vec{A}|^6} + \ldots \right) \ .
  \ee
This does not look similar to 
 \be
\label{corr31}
{\cal L}_{\rm eff}^{(3+1)} \ =\ -\frac {F_{\mu\nu}^2}{4e^2(\mu)} 
\ =\ 
 -\frac {F_{\mu\nu}^2}4 \left[ \frac 1{e_0^2} + 
\frac 1{4\pi^2} \ln \frac {\Lambda}{\mu} 
+ \frac {e_0^2}{16\pi^4} \ln \frac {\Lambda}{\mu} + \ldots \right]
 \ .
  \ee
As we will see, there is also no obvious correspondence between the
 individual
cotributions in  ${\cal L}_{\rm eff}^{(0+1)}$ and  
${\cal L}_{\rm eff}^{(3+1)}$.
To be precise, the diagrams determining the corrections in these 
two cases are
identical, but the results of their evaluation are not similar. 

We still believe that the correspondence between higher order 
corrections to
${\cal L}_{\rm eff}$ in different dimensions will eventually be 
unravelled, but at the moment we 
can only
report our accurate calculation of the term $ - 3/(4|\vec{A}|^6)$ 
in 
Eq.(\ref{corr01}). 

Before proceeding with it, let us make a remark on the {\it title} 
of our
paper. The term ``matrix models'' usually refers  to quantum-mechanical 
versions of 
{\it non-Abelian} gauge theories (the dynamic variables in such 
theories
are matrices). Matrix models (in the first place, maximally 
supersymmetric
matrix models obtained by dimensional reduction of 4D ${\cal N} =4$
theories) attracted recently a considerable attention due to their 
implications for strings/branes/M--theory dynamics. In particular, 
the 
corrections $\propto \dot{\vec{A}}^2$ to the effective Lagrangian 
in the
maximally supersymmetric matrix models vanish, while the nontrivial
corrections  $\propto (\dot{\vec{A}}^2)^2$ and of still higher 
order in derivatives 
can be related to the scattering amplitudes of gravitons in 
11--dimensional 
space \cite{Duglas}-\cite{Plefka}.

In the Abelian models discussed in this paper, gauge and fermion 
fields
do not have matrix form. In addition, these models probably do not 
have stringy or gravity implications (our primary interest here is 
the effective Lagrangian itself). But their kinship to non--Abelian
 matrix models is obvious.

\section{Calculational set-up}
\setcounter{equation}0

Supersymmetric QED involves the gauge field $A_\mu$, the 
photino (Majorana fermion) 
field $\lambda$, two charged scalars $\phi$, $\chi$, and a 
charged Dirac fermion $\psi$. We assume that the charged 
fields are massless. In the gauge $A_0 = 0$, 
the Lagrangian of the dimensionally reduced 
theory has the form
  \be
\label{LSQED}
{\cal L}\ =\ \frac 12 \dot{A_k}^2 + \dot{\varphi} \dot{\bar\varphi}
 + 
\dot{\chi} \dot{\bar\chi} + i[\bar \lambda \dot{\lambda} + 
\bar \psi \dot{\psi}]  \nonumber \\
- e^2 (\bar\varphi \varphi + \bar\chi \chi) A_k^2 
  - \frac 12 e^2 (\bar \varphi \varphi - \bar\chi \chi)^2 
 +i eA_k \bar \psi \gamma_k \psi + \nonumber \\
 e\sqrt{2}  \left[  \chi  \bar  \psi_L \lambda + \bar \chi \bar \lambda \psi_L
- \phi  \bar  \psi_R \lambda - \bar \phi \bar \lambda \psi_R 
 \right]
  \ee
[$\psi_{L,R} = \frac 12 (1 \mp \gamma^5) \psi$; 
$\gamma_k$ are Euclidean, i.e. Hermitian. We will set in what follows
 $e \equiv 1$].  
Consider the system in a 
constant gauge field background $\vec{A}(t) = \vec{C}$ [in (0+1) 
dimensions, not only the field strength $\vec{E} = \dot{\vec{A}}$, but
 also the potential
$\vec{A}$ cannot be disposed of by a gauge transformation and has 
direct physical meaning]. Then the charged fields $\phi, \chi, \psi$ 
acquire the mass $|\vec{A}|$. If this mass is much larger 
than the energy scale set up by the gauge kinetic term 
$(1/2)\partial^2/(\partial \vec{A})^2$ in the effective 
Hamiltonian, 
$E_{\rm char} \sim 1/\vec{A}^2$, we can treate the charged 
variables as ``fast" (in the Born--Oppenheimer framework) and 
integrate them over. Actually, in a constant background all 
corrections to the effective {\it potential} vanish, this is 
guaranteed by supersymmetry. 

Nontrivial corrections to ${\cal L}_{\rm eff}$ are obtained if 
considering a slowly changing background 
  \be
  \label{fon}
  \vec{A}(t) \ =\ \vec{C} + \vec{E}t \ .
   \ee
 The leading correction (the one which vanishes in the 
${\cal N} = 4$ case) is proportional to $\vec{E}^2$.  The 
calculation of the 1--loop contributions to ${\cal L}_{\rm eff}$ 
was described in \cite{Akh}, and
our task here is to tackle the two-loop graphs drawn in Fig. 1 in 
the background (\ref{fon}). 
\begin{figure}
   \begin{center}
        \epsfxsize=300pt
        \epsfysize=200pt
        \vspace{-5mm}
        \parbox{\epsfxsize}{\epsffile{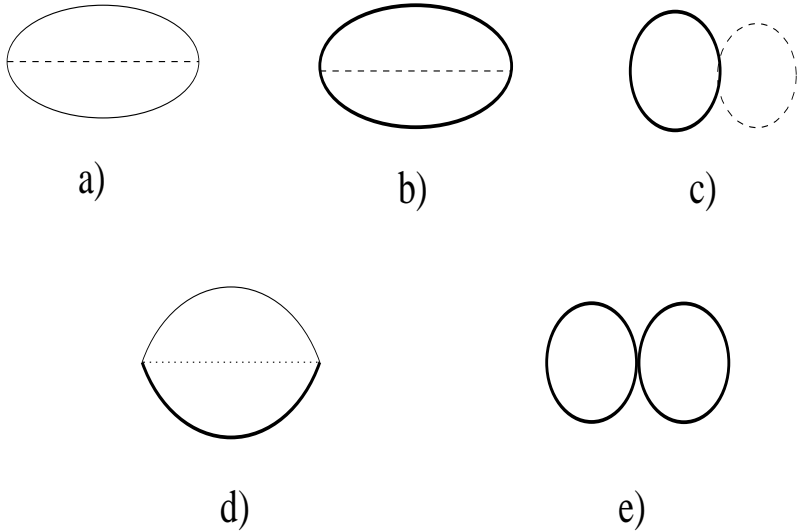}}
        \vspace{-5mm}
    \end{center}
\caption{Graphs contributing to ${\cal L}_{\rm eff}^{(2)}$. Thin 
solid lines describe fermions, bold lines -- scalars, dashed lines
 - photon and dotted line - photino. }
\label{dvepetli}
\end{figure}
 
 The calculation to be done is very similar in spirit to the 
calculation of charge renormalization in 4D SQED performed in
Ref. 
\cite{VShif}. In 4 dimensions, one also has to evaluate the graphs 
in Fig. 1 substituting there the Green's functions in the constant 
field strength background
\footnote{The corresponding technique was developped in 
Ref.\cite{fpg}.}
 \be
 \label{fon4}
 A_\mu \ =\ - \frac 12 F_{\mu\nu} x_\nu\ .
  \ee
  Adding all contributions, one obtains
  \be
  \label{beta2}
  \beta^{(2)} \ =\ \frac 1{\pi^2}
  \ee
  for the second coefficient in the $\beta$ function, which leads 
to (\ref{corr31}). 

The calculation in the QM limit is much more difficult, however, 
because {\it (i)} Lorentz invariance is lost and {\it (ii)} in 
contrast to the background (\ref{fon4}), the background (\ref{fon})
 is {\it genuinely} noninvariant with respect to time translation. 
A similar
(actually, more complicated) calculation was performed 
 for non-Abelian theories\cite{sestry,Okawa}. Unfortunately, these 
papers are not written in a "user-friendly" way and, for the benefit
 of future users, we took pain to describe the technical details of
 the calculation at some length.
  
\section{Boring technicalities.}
\setcounter{equation}0

First of all, we go over into Euclidean space, $t \to -i\tau$. The 
background  is still 
  \be
\label{fonE}
\vec{A}(\tau)  = \vec{C } + \vec{E}\tau
\ee 
with Euclidean $\vec{E}_E$ which differs from the physical Minkowskian
$\vec{E}_M$ by the factor $i$. We will calculate the effective action in the
Euclidean background (\ref{fonE}) assuming $\vec{E}_E$ real, not forgetting
to change the sign of the term $\propto \vec{E}^2$ in the end of the day.
 In the background (\ref{fon}), the scalar and fermion 
Green's functions $D$ and $G$ depend on both initial and final times
$\tau, \tau'$ in a nontrivial way. The exact expressions can be 
found in Refs. \cite{sestry,Okawa}. Instead of working with them 
directly, it is
convenient to make a Fourrier transform over the variable 
$\tau_- = \tau' - \tau$ and represent 
  \be
  \label{DGsym}
  D(\tau, \tau') &=& \int \frac {d\epsilon}{2\pi} 
e^{i\epsilon\tau_-} D(\epsilon, \tau_+) \ , \nonumber \\
 G(\tau, \tau') &=& \int \frac {d\epsilon}{2\pi} 
e^{i\epsilon \tau_-} G(\epsilon, \tau_+) \ ,
   \ee
where $\tau_+ = (\tau + \tau')/2$. Next, we expand $D(\epsilon, 
\tau_+) $ and $G(\epsilon, \tau_+) $ in  $\vec{E}$ at fixed
$\vec{A}(\tau_+) = \vec{C} + \vec{E}\tau_+$. 
The leading terms are 
 \be
 \label{DGlead}
 D^{(0)}(\epsilon, \tau_+) &=& \frac 1{\epsilon^2 + 
\vec{A}^2(\tau_+)} \ , \nonumber \\
G^{(0)}(\epsilon, \tau_+) &=& \frac 1{\epsilon \gamma^0 + 
\vec{A}(\tau_+) \vecg{\gamma} }\ ,
 \ee
 where $\gamma^0$ and $\vecg{\gamma}$ are Euclidean $\gamma$ 
matrices, $\gamma_\mu \gamma_\nu + 
 \gamma_\nu \gamma_\mu  = 2\delta_{\mu\nu}$. For our purposes, 
we only  need linear and quadratic terms in the expansion of
$D(\epsilon, \tau_+)$, $G(\epsilon, \tau_+) $ in $\vec{E}$. One can 
work them out from the general expressions of Refs.
 \cite{sestry,Okawa} or,
alternatively, use the results of Shuryak and Vainshtein 
\cite{VShur} who calculated the 
4D scalar and fermion Green's functions in a generic Euclidean 
constant field strength background,
\footnote{If performing the average $F_{\alpha\beta} 
F_{\beta\gamma} \to -\frac14 \delta_{\alpha\gamma}\langle F^2 
\rangle$ as is usually done in most (though not all, hence the 
paper \cite{VShur}) QCD applications, 
the quadratic terms in Eq. (\ref{DGVShur}) vanish. }
  \be
  \label{DGVShur}
  D(p) &=& \frac 1{p^2} - \frac 2{p^8} \left[ p_\alpha 
F_{\alpha\beta} F_{\beta\gamma} p_\gamma + \frac 14 F_{\mu\nu}^2 
p^2 \right]\ , \nonumber \\
G(p) &=& \frac 1{/\!\!\!p} + \frac {iF_{\alpha\beta} }{4p^4}
(/\!\!\!p \sigma_{\alpha\beta} + \sigma_{\alpha\beta} /\!\!\!p )
- \frac {2p_\alpha}{p^8}F_{\alpha\beta} F_{\beta\gamma} 
(/\!\!\!p  p_\gamma - p^2 \gamma_\gamma) \ .
  \ee
In our case, we have to set $p = (\epsilon, \vec{A}), 
F_{0j} =   E_j, 
F_{jk} = 0$. We obtain
   \be
   \label{DGexpan}
 D(\epsilon, \tau_+) &=& \frac 1{p^2} + \frac 1{p^8}\left[ 
 \vec{E}^2 (\epsilon^2 - \vec{A}^2) + 2(\vec{E} \vec{A})^2 \right] 
\ , \nonumber \\
G(\epsilon, \tau_+) &=& \frac 1{/\!\!\!p} + \frac 
{iE_k}{2p^4}(/\!\!\!p \gamma^0 \gamma^k + 
\gamma^0 \gamma^k /\!\!\!p ) \nonumber \\
&+& \frac 2{p^8}\left[ \epsilon \vec{E}^2 (\epsilon 
/\!\!\!p  - p^2 \gamma^0) + /\!\!\!p (\vec{E} \vec{A})^2 - p^2 
(\vec{E} \vec{A})(\vec{E} \vecg{\gamma}) \right]
 \ee
 ($p^2 \equiv \epsilon^2 + \vec{A}^2, \ /\!\!\!p  \equiv \epsilon 
\gamma^0 + \vec{A} \vecg{\gamma}$ and $\vec{A}$ is evaluated at 
$\tau_+$). 
 The photon Green's function can be chosen in the gauge $A_0 = 0$
 (which is equivalent to the Landau gauge $\partial_\mu A_\mu = 0$ in 
the QM limit),
   \be
   \label{photon}
   D_{jk}(\tau - \tau') \ =\ 
    \delta_{jk} \int  \frac {d\omega }{2\pi \omega^2} e^{i\omega 
(\tau - \tau')} \ .
  \ee 
  The 3--point scalar--photon vertex entering the graph in 
Fig. 1b appears due to nonvanishing background :
 \be
 \label{scalphot}
\Gamma_{\bar\phi \phi A_k} 
 = \Gamma_{\bar\chi \chi A_k} 
\ =\ 
-2i(C_k + E_k\tau) \ .
 \ee
 The graphs in Fig. 1 determine the correction to the effective 
action $S_{\rm eff}$. The corresponding analytic expressions involve
 the integrals over the time moments referring to the vertices --- 
the integral $\int d\tau$ for the figure-eight graphs in 
Fig. 1c,e and the double
integral over $d\tau d\tau'$ for other graphs. Let us represent
$d\tau d\tau' = d\tau_- d\tau_+$ and {\it postpone} the integration
 over $d\tau_+$ (over $d\tau$ for the fugure-eight graphs). The
effective action is expressed as 
 \be
 \label{pseudoL}
 S_{\rm eff} \ =\ \int d\tau \tilde{\cal L}(\tau) \ .
   \ee
   Let us call the integrand $ \tilde{\cal L}(\tau) $ 
{\it pseudo-Lagrangian}.  The pseudo-Lagrangian  depends on $\tau$ 
only via $\vec{A}(\tau)$.  $ \tilde{\cal L} $ can be calculated 
using standard Feynman rules. One only has to take into account 
the modification of the charged propagators as written in 
Eq.(\ref{DGexpan}) and the appearance of the new vertex 
(\ref{scalphot}). We obtain
  \be
  \label{tildeL}
  \tilde{\cal L} \ =\ - \frac 12 \int \frac {d\epsilon}{2\pi} 
\int \frac 
  {d\omega}{2\pi\omega^2} {\rm Tr} \{\gamma_j G(\epsilon)
  \gamma_j G(\epsilon + \omega) \} + \nonumber \\
   \int \frac {d\epsilon}{2\pi} \int \frac 
  {d\omega}{2\pi\omega^2} \left[ \vec{A}^2 
D(\epsilon) D(\epsilon + \omega)
  + \vec{E}^2 D''(\epsilon) D(\epsilon + \omega) \right] -
\nonumber \\
  6 \int \frac {d\epsilon}{2\pi} D(\epsilon) \int \frac 
  {d\omega}{2\pi\omega^2} 
  - 2\int \frac {d\epsilon}{2\pi} \int \frac 
  {d\omega}{2\pi\omega}  {\rm Tr} \{\gamma^0 G(\epsilon) \}
 D(\epsilon + \omega) \nonumber \\  - \left[ \int \frac {d\epsilon}{2\pi}  
D(\epsilon) \right]^2\ ,\ \ \ \ \ \ \ \ \ \ 
  \ee
  where the five terms above correspond to the 
five graphs in Fig. 1.  
In the case of constant external field, different contributions in 
Eq.(\ref{tildeL}) exactly cancel each other as they should. 
 
 When $\vec{E} \neq 0$, $\tilde{\cal L}$ does not vanish. Note 
first of all that certain terms in Eq.(\ref{tildeL}) involve a 
power infrared divergence
 \be
 \label{infrmu}
\int \frac   {d\omega}{2\pi\omega^2} \to  \int \frac 
{d\omega}{2\pi(\omega^2 + \mu^2) }  \
 =\ \frac 1{2\mu}
  \ee
  ($\mu$ is a fictitious photon mass). This divergence vanishes 
as everything else does when the background is constant, but the 
divergence 
{\it survives} when $\vec{E} \neq 0$. This divergences have 
actually a rather transparent physical meaning which will be 
clarified a little bit
later. For the time being let us just {\it ignore} the infrared 
divergent terms and present the result of our calculation for the 
{\it finite} pieces in  $\tilde{\cal L}$.

The fermion loop   in Fig. 1a gives
\be
\label{1a}
\tilde{\cal L}_{1a} \ =\ \frac {7[\vec{E}^2 A^2 - (\vec{E} 
\vec{A})^2]}{16 A^8}
 \ee
 ($A = |\vec{A}|$).
 The scalar loop of Fig. 1b gives
  \be
  \label{1b}
  \tilde{\cal L}_{1b} \ =\ \frac {\vec{E}^2}{2A^6} - 
  \frac {7(\vec{E} \vec{A})^2}{8A^8}\ .
  \ee
The graph in Fig. 1c does not give a finite contribution 
(only an infrared divergent one). The photino graph in Fig. 1d 
gives
\be
  \label{1d}
  \tilde{\cal L}_{1d} \ =\ -\frac {5\vec{E}^2}{8A^6} + 
  \frac {5(\vec{E} \vec{A})^2}{4A^8}\ .
  \ee
And finally the scalar self-interaction graph in Fig. 1e gives
\be
  \label{1e}
  \tilde{\cal L}_{1e} \ =\ \frac {\vec{E}^2}{8A^6} - 
  \frac {5(\vec{E} \vec{A})^2}{16A^8}\ .
  \ee
Adding all pieces, we obtain
 \be
  \label{tildeLres}
  \tilde{\cal L}^{\rm 2\ loops} \ =\ \frac {7\vec{E}^2}{16A^6} - 
  \frac {3(\vec{E} \vec{A})^2}{8A^8}\ .
  \ee
This cannot be the correct result for the effective  Lagrangian, 
however. As was mentioned before, supersymmetry requires  the 
metric to have a conformally flat form and the terms $\propto 
(\vec{E} \vec{A})^2$ are not allowed. The paradox is solved if 
noting
that by calculating the graphs in time-dependent background we 
{\it can}not calculate directly the effective Lagrangian, but only 
the effective action which is equal to the integral of the 
pseudo-Lagrangian 
$\tilde{\cal L}$ according to the definition (\ref{pseudoL}). 
When doing this integral, we have to restore the time-dependence 
in Eq.(\ref{tildeLres}). It is convenient (though not at all 
necessary) to assume  that the constant and linear term are 
orthogonal to each other, $\vec{C} \vec{E} = 0$. Then
$$A^2 \to C^2 + E^2 \tau^2, \ \ \ \ \ \ \ \ 
\vec{E} \vec{A} \to E^2 \tau \ .$$
The integrals are easily done:
 \be
 \label{inttau}
 \int_{-\infty}^\infty d\tau  \frac {E^2}{(C^2 + E^2 \tau^2)^3} \ 
=\ \frac {3\pi E}{8C^5}\ , \nonumber \\
\int_{-\infty}^\infty d\tau  \frac {E^4 \tau^2}{(C^2 + E^2 
\tau^2)^4} \ =\ \frac {\pi E}{16C^5}
 \ee
 In other words, the contribution of the structure $(\vec{E} 
\vec{A})^2/A^8 $ in $\tilde{\cal L}$ into $S_{\rm eff}$ is exactly 
6 times less than the contribution of the structure 
$\vec{E}^2/{A}^6$. To find the effective Lagrangian, we 
should take into account the requirement of supersymmetry 
${\cal L} \propto
\vec{E}^2/{A}^6$ and also require that 
$S_{\rm eff} \ =\ \int d\tau {\cal L}_{\rm eff}(\tau)$. 
This finally gives
  \be
\label{L2loop}
 {\cal L}_{\rm eff}^{\rm 2 \ loops} \ =\  \frac {3\vec{E}_E^2}{8A^6}
\ =\  -\frac {3\vec{E}_M^2}{8A^6}
\ ,
 \ee
from which the result (\ref{corr01}) follows.

Exactly this procedure [{\it (i)} calculating the full time
integral
for $S_{\rm eff}$ and {\it (ii)} restoring from this 
${\cal L}_{\rm eff}(\tau)$ using  supersymmetry 
requirements ] was used (if not spelled out explicitly) 
in Ref.\cite{sestry}. We convinced ourselves in it after 
partially reproducing their calculations (see Appendix). 

The situation may seem somewhat paradoxical. After all, the
effective Lagrangian has a well defined meaning: one can
derive it by canonical rules from the effective 
Born--Oppenheimer Hamiltonian calculated using the philosophy
of Ref.\cite{jaSQED}
\footnote{We note, however, that such calculation is going
to be rather difficult. The contribution 
$\sim \vec{E}^2/A^6$ in $H_{\rm eff}$ involves two extra
orders in the Born--Oppenheimer parameter 
$\gamma \sim x^{\rm fast}/x^{\rm slow} \sim 1/\sqrt{A^3}$
compared to that calculated in \cite{jaSQED}.}  
  On the other hand, we cannot calculate it {\it directly} in the
Lagrangian approach. In particular, one can ask what happens 
in the nonsupersymmetric case where both structures 
$\vec{E}^2/A^6$ and $(\vec{E} \vec{A})^2/A^8$ are allowed. 
The answer to the
last question is clear, however. The matter is that, 
in  nonsupersymmetric case, one just 
cannot consistently define what {\it is} an effective Lagrangian 
(or Hamiltonian). Indeed, the zero point energy contribution
to  a nonsupersymmetric
effective potential is $\sim A$ which is of the same order as
the characteristic energy of the fast charged field
 excitations and the necessary separation of scales is absent.

On the contrary, in supersymmetric case the notion of 
$H_{\rm eff}$ is well defined. At the same time, supersymmetry
imposes constraints on the allowed form of the effective 
Lagrangian which is thereby restored unambiguously. 
If you will, you
may consider this as an {\it indirect  proof} that the metric on 
supersymmetric moduli space should be conformally flat.

Let us return now to the question of infrared divergences. They
are present in the graphs in Fig. 1(a,b,c). Introducing the photon
mass as in Eq.(\ref{infrmu}), we derive
 \be
\label{infrabc}
\tilde {\cal L}^{IR}_{1a} \ =\ 
 \frac {5[ (\vec{E} 
\vec{A})^2 - \vec{E}^2 A^2 ]}{16  \mu A^7}\ , \nonumber \\
\tilde {\cal L}^{IR}_{1b} \ =\ 
 \frac {35(\vec{E} 
\vec{A})^2 - 17\vec{E}^2 A^2}{32  \mu A^7} \nonumber \\
\tilde {\cal L}^{IR}_{1c} \ =\ 
\frac {3\vec{E}^2}{8\mu A^5} - 
\frac {15(\vec{E} 
\vec{A})^2}{16  \mu A^7} 
   \ee
and
 \be
\label{infrL}
\tilde {\cal L}^{IR}_{\rm tot} \ =\ 
 \frac {15[(\vec{E} 
\vec{A})^2 - \vec{E}^2 A^2]}{32  \mu A^7}\
   \ee
for the sum. Integrating this over $d\tau$, we obtain
 $$  -\frac { E}{2  \mu C^4} \ .$$
Here the contribution of the term $(\vec{E} \vec{A})^2$ in the
integral is {\it five} (rather than six) times smaller than the
contribution of the term $\vec{E}^2 A^2$. This gives
 \be
\label{Linfr}
   {\cal L}^{IR} \ =\ 
  \frac {3\vec{E}^2}{ 8\mu A^5} 
  \ee
in Minkowski space.
It is clear, however, that the divergence is due to the virtual
photons with low energies while the true effective Lagrangian
is related to the graphs with only heavy degrees of freedom
(with energy $\sim A$) circulating in the loops. The contribution
(\ref{Linfr}) should be actually interpreted as the iteration
of ${\cal L}_{\rm eff}^{\rm 1\ loop}$, as shown in Fig. 2, and 
should not be included into the definition of  
  ${\cal L}_{\rm eff}^{\rm 2\ loops}$. Indeed, evaluating the 
diagram in Fig. 2, we obtain
\be
\Delta  {\cal L}^{IR}  =  \frac 12 \left[
\frac {\partial^2}{\partial A_j \partial A_j } 
{\cal L}_{\rm eff}^{\rm 1\ loop} \right] 
\int \frac {id\omega_M} {2\pi \omega_M^2}  = 
\frac {\vec{E}^2}2
\Delta \left( \frac 1{4A^3} \right) 
 \int \frac {d\omega_E }{2\pi \omega_E^2}
 \ee
($\omega_M$ and $\omega_E$ are the photon frequencies before and
after Wick's rotation), which coincides with Eq. (\ref{Linfr}).

\begin{figure}
   \begin{center}
        \epsfxsize=150pt
        \epsfysize=80pt
        \vspace{-5mm}
        \parbox{\epsfxsize}{\epsffile{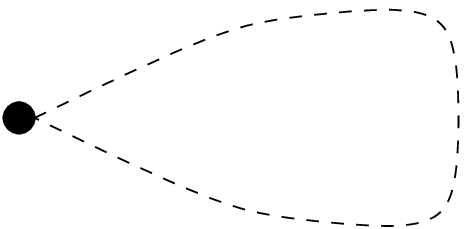}}
        \vspace{-5mm}
    \end{center}
\caption{Iteration of ${\cal L}_{\rm eff}^{1\ loop}$.}
\label{itera}
\end{figure}

Note that the infrared divergencies of the discussed type
cancel out in the two--loop contributions to the quartic terms
$\propto (\vec{E}^2)^2 $ in ${\cal L}_{\rm eff}$ in ${\cal N} = 4$
theories \cite{Dine}. This simply follows from the fact that the
one--loop contribution in  ${\cal L}_{\rm eff}$ is proportional
to $ (\vec{E}^2)^2/A^7 $, where $\vec{A}$ is now a 9--dimensional 
vector and that the 9--dimensional Laplacian of $1/A^7$ vanishes
for $A \neq 0$. In the ${\cal N} =1$ case, the infrared 
divergences survive in the sum of the contributions in Fig. 1, but,
as we explained, they anyway should not be taken into account in
  ${\cal L}_{\rm eff}$. 

\section{${\cal N} = 2$ Electrodynamics}
\setcounter{equation}0

This calculation can be easily generalized to the case 
$4D$, ${\cal N} =2$ QED, which is equivalent to  
$6D$, ${\cal N} =1$ QED in the quantum--mechanical limit.
Thinking in the (3+1)-dimensional terms, we have now two 
(rather than one) photino fields
$\lambda_{1,2}$  and  two extra real neutral scalars $a$ and $b$
[in the (5+1)--dimensional language they correspond to the components
$A_{4,5}$ of the gauge field].
 As in the ${\cal N} = 1$ case, we have a Dirac 
spinor and two complex scalars in the charged sector. 

The form of the reduced Lagrangian can be easily derived from the
known 4--dimensional expression \cite{Fayet}. We have
   \be
\label{LSQEDN2}
{\cal L}\ =\ \frac 12 \left(\dot{A_k}^2 + \dot a^2 + \dot b^2 \right)
+ \dot{\varphi} 
\dot{\bar\varphi} + 
\dot{\chi} \dot{\bar\chi} + i\sum_{f=1,2} 
\bar \lambda_f \dot{\lambda}_f + 
i \bar \psi \dot{\psi} \nonumber \\ 
-  (\bar\varphi \varphi + \bar\chi \chi) \left( A_k^2 + a^2 + b^2 \right) 
  - \frac 12  (\bar \varphi \varphi + \bar\chi \chi)^2 
 +i A_k \bar \psi \gamma_k \psi 
+ \bar\psi(a + ib\gamma^5) \psi 
\nonumber \\ 
+  \sqrt{2}  \left[ \chi (\bar  \psi_L  \lambda_1  + \bar \psi_R  \lambda_2 )
-   \phi (\bar  \psi_R  \lambda_1  + \bar \psi_L  \lambda_2 ) \ + {\rm H.c.}
\right] \ .\ \ \ \ \ \ 
  \ee
The effective Lagrangian described the motion over 5--dimensional 
moduli space $(A_k, a, b)$.
Let us calculate it assuming as before that the background has
the form (\ref{fon})  (i.e.  $a$ and $b$  vanish).
This is convenient because we can use the same expressions 
(\ref{DGexpan}) for the Green's functions as before. 
For a generic background, 
${\cal L}_{\rm eff}$ can be restored from rotational $O(5)$ invariance
and supersymmetry.

The effective action is described as before by the graphs in 
Fig. 1, where the dashed lines stand now for the gauge fields and also 
the
neutral scalars $a,b$.  It is obvious that the photino contribution of 
Fig. 1d
is now multiplied by 2. A simple combinatorics tells us that 
the contribution due to scalar 
self--interactions of Fig. 1e is multiplied by the factor 3
compared to the ${\cal N} =1$ case due to the 
{\it positive} relative sign in 
 the scalar potential in Eq.(\ref{LSQEDN2}).
The contribution of the scalar loop in Fig.1b is the same as 
before (there is no contribution from $a,b$ exchange because the background 
was chosen 3--dimensional). The contribution of the 
graph in Fig.1c is multiplied by $5/3$ (it is just the counting
of degrees of freedom). Extra
degrees of freedom $a,b$ also give  
a nontrivial contribution in the fermion loop in Fig. 1a,
 \be
\label{delferm}
\Delta \tilde {\cal L}_{\rm ferm} \ =\ 
\int \frac {d\epsilon}{2\pi} \int \frac {d\omega}{2\pi \omega^2}
{\rm Tr} \left\{ G(\epsilon) G(\epsilon + \omega) \right\}
\nonumber \\
= \  - \frac {[\vec{E}^2 A^2 - (\vec{E} 
\vec{A})^2]}{8 A^8} +  \frac {5[\vec{E}^2 A^2 - (\vec{E} 
\vec{A})^2]}{16 \mu A^7}
\ ,
  \ee   
The sum of all infrared divergent pieces gives 
 \be
 \frac {3\vec{E}^2}{32\mu A^5} -  \frac {15 (\vec{E} \vec{A})^2}
{32 \mu A^7}
 \ee
As was explained in the previous section, we have to substitute here
$(\vec{E} \vec{A})^2/A^7 \to \vec{E}^2 /(5A^5)$, after which the 
contribution vanishes. This cancellation has the same origin as
the cancellation of the infrared divergences $\propto (\vec{E}^2)^2/
(\mu A^9)$ in ${\cal N} = 4$ theory mentioned above
 and follows from the fact that the 5--dimensional Laplacian of
$1/(\vec{A}^2+a^2+b^2)^{3/2}$ vanishes.  

The finite contribution to the pseudo-Lagrangian is
 \be
\label{TilLN2}
\tilde{\cal L}^{\rm 2\ loops}_{{\cal N} =2}(\tau) 
\ =\ -\frac {\vec{E}^2}{16A^6} + 
  \frac {3(\vec{E} \vec{A})^2}{8A^8}\ .
 \ee 
The coefficient of the second term in Eq.(\ref{TilLN2}) is exactly
6 times greater than that for the first term and 
the integral of this expression over $d\tau$, from which the 
effective Lagrangian is extracted, just {\it vanishes}. 
This should have been expected, of course: supersymmetry
and rotational invariance
require that all higher--loop corrections to the effective
Lagrangian of  ${\cal N} = 2$ theory vanish \cite{Entin}.

\section{Conclusions}
The main result of this paper is quoted in the abstract. We also verified that
the two--loop corrections vanish in the $ {\cal N} =2$ case. We did not
achieve our goal, however, and did not establish an operative
 relationship between the two--loop corrections to ${\cal L}_{\rm eff}$ 
in the matrix models and in $4D$ theories. For example, the figure-eight
graph gives a nontrivial contribution (\ref{1e}) in the $(0+1)$ case, but 
its contribution to the $4D$ $\beta$ function vanishes (this follows from 
the vanishing of the corrections to the scalar Green's function in the 
Lorentz--invariant situation).  

On the other hand, one can note that
the 4--dimensional $\beta$ function and effective Lagrangian are not,
strictly speaking, uniquely defined beyond one loop. There is
the Wilsonean definition, according to which  the higher loops
vanish not only for
${\cal N} =2$, but also for ${\cal N} =1$ theories.  
Eq.(\ref{corr31})  is written in the conventional definition where $\beta$
function involves the large-distance contribution associated with
anomalous dimensions of the charged fields. \cite{VShif,NVZS}. Perhaps,
the metric in Eq.(\ref{corr01}) (defined quite unambigously)  is related
to the 4--dimensional $\beta$ function defined in some particular
physically relevant way ?
Further studies in this direction are welcome.

I am  indebted to K. Becker, 
J. Plefka, M. Shifman, and A. Waldron
 for useful discussions and correspondence. 

\section*{Appendix. A sample non-Abelian calculation.}
\setcounter{equation}0
 \renewcommand{\theequation}{A.\arabic{equation}}

For pedagogical purposes, we sketch here a sample  calculation of the 
``figure-eight'' gauge boson graph for the $SU(2)$
${\cal N} =4$ matrix model in 
some more details than it was done in Ref.\cite{sestry}. 

Choose the Abelian
background $A_0^{\rm cl} = 0$ and
   \be
  \label{fonN4}
   \vec{A}^{\rm cl}(\tau) \ =\ \frac 12(\vec{b} + \vec{v}\tau)\sigma^3 \ ,
   \ee
where $\vec{b}$ and $\vec{v}$ are two orthogonal 9--dimensional vectors. 
We are using the notations of Ref.\cite{sestry} (
$\vec{r} = \vec{b} + \vec{v} \tau $ is interpreted as the distance between 
two $D0$ particles, $\vec{b}$ is their impact parameter, and 
$\vec{v}$ is their respective velocity), bearing
in mind the identifications
 $\vec{r} \equiv \vec{C}, \vec{v} \equiv \vec{E}_E$. 

We
represent $ {\cal A}_\mu^{\rm qu} =  {A}_\mu^{\rm cl} + a_\mu$ and 
impose the
{\it Feynman background gauge} \cite{background} adding to the Lagrangian 
the term 
$-\frac 12 ({\cal D}_\mu a_\mu )^2$, where ${\cal D}_\mu$ is the covariant
derivative associated with the background. We obtain for the quadratic and 
quartic part of the gauge field Euclidean Lagrangian
 \be
\label{partL}
{\cal L} \ =\ - {\rm Tr} \left\{ a_\mu \left( {\cal D}^2 a_\mu - 2i 
[F_{\mu\nu}^{\rm cl}, a_\nu] \right) \right\}  - \frac 12 
 {\rm Tr} \left\{ [a_\mu, a_\nu]^2  \right\} + {\rm other\ terms}
 \ee
The Abelian fluctuation components   $a_\mu^{\rm Ab} \propto \sigma^3$ 
do not interact with the
background and remain massless. The corresponding fugure-eight graphs 
are infrared divergent (like the graph in Fig. 1c above). Following
Ref.\cite{sestry}, we ignore them here. On the other hand, the components
$\propto \sigma^{1,2}$ provide a nontrivial finite contribution 
in ${\cal L}_{\rm eff}$. It is convenient to represent
$a_\mu = (\bar \phi_\mu \sigma^+ + \phi_\mu \sigma^-)/\sqrt{2}$, after which
 Eq.(\ref{partL}) is rewritten as 
 \be
\label{partLphi} 
 \dot{\bar \phi_\mu}  \dot\phi_\mu + (\vec{b}^2 + \vec{v}^2 \tau^2)
 \bar \phi_\mu \phi_\mu
- 2i F_{\mu\nu}  \bar \phi_\mu \phi_\nu - \frac 12 ( \bar \phi_\mu \phi_\nu 
-  \bar \phi_\nu \phi_\mu)^2
 \ee
($F_{0j} = v_j,\ F_{jk} = 0$). The contribution of the figure-eight graph to  
the effective pseudo-Lagrangian
$\tilde {\cal L}_{\rm eff}$ has the form
 \be
\label{Delmunu}
\Delta_{\mu\nu} \Delta_{\mu\nu} - \frac 12 \Delta_{\mu\nu} \Delta_{\nu\mu}
 - \frac 12 \Delta_{\mu\mu} \Delta_{\nu\nu} \ ,
\ee
where $\Delta_{\mu\nu}$ is the Green's function 
$\langle \bar \phi_\mu(\tau)  \phi_\nu(\tau) \rangle$. 
Now, if the term $\propto F_{\mu\nu}$ in Eq.(\ref{partLphi}) 
were absent, $\Delta_{\mu\nu}$
 would exactly coincide with 
the scalar Green's function in Eqs.(\ref{DGsym}),
(\ref{DGexpan}), multiplied by $\delta_{\mu\nu}$. In this  case, the 
quadratic in $E$ terms in the pseudo-Lagrangian would have the same form
as in Eq.(\ref{1e}) up to the overall factor $\left[(\delta_{\mu\mu})^2 - 
\delta_{\mu\nu} \delta_{\mu\nu}\right]/2 = 45$. Doing the integral 
over $d\tau$ and
assuming that $\int d\tau \tilde {\cal L}_{\rm eff}(\tau)  = 
\int d\tau {\cal L}_{\rm eff}(\tau)$ and that 
the
effective Lagrangian does not involve the terms $\propto (\vec{v} \vec{r})^2$,
this gives the correction 
 \be
\label{105}
\frac {105 \vec{v}^2}{32r^6} 
 \ee
in  ${\cal L}_{\rm eff}$. Let us now take into account the modification
of $\Delta_{\mu\nu}$ due to the presence of the
term $\propto F_{\mu\nu}$ in the quadratic part of the Lagrangian. As we are
interested only in the terms $\propto \vec{v}^2$, we can neglect 
now 
$\vec{v}$--dependent
terms in ${\cal D}^2$ (or, better to say, treat $r^2 = \vec{b}^2 +
\vec{v}^2 \tau^2$ as a constant).  We obtain
  \be
\Delta_{\mu\nu}\ \sim \ \int \frac {d\omega}{2\pi} \| (\omega^2 + r^2) 
\delta_{\mu\nu} - 2i F_{\mu\nu} \|^{-1} \ = \nonumber \\
\int \frac {d\omega}{2\pi} \left[ \frac {\delta_{\mu\nu}}{\omega^2 + r^2}
+ \frac { 2i F_{\mu\nu}}{(\omega^2 + r^2)^2}- \frac 
{4F_{\mu\alpha}F_{\alpha\nu}}{(\omega^2 + r^2)^3}  + \ldots \right]\ .
  \ee
Substituting this into (\ref{Delmunu}) and adding the term (\ref{105}),
we obtain
\be 
  - \frac {45}{4r^2} - \frac{15\vec{v}^2}{2r^6} + \frac {105 \vec{v}^2}{32r^6}
\ =\  - \frac {45}{4r^2} - \frac {135 \vec{v}^2}{32r^6} \ ,
  \ee
which coincides with the sum of the contributions (5.6) and (5.7) in
Ref.\cite{sestry} (with  the terms $\propto \vec{v}^4/r^{10}$ neglected).

\end{document}